\documentclass[draft,jgrga]{AGUTeX}
\usepackage{graphicx}
\usepackage{amsfonts}
\usepackage{color}
\authorrunninghead{XIONG ET AL.}
\titlerunninghead{MC--Shock Interaction and Its Geoeffectiveness}

\authoraddr{Ming Xiong, Huinan Zheng, Yuming Wang, and Shui Wang,
CAS Key Laboratory for Basic Plasma Physics, School of Earth and
Space Sciences, University of Science and Technology of China,
Hefei, Anhui 230026, China (mxiong@mail.ustc.edu.cn;
hue@ustc.edu.cn; ymwang@ustc.edu.cn; and swan@ustc.edu.cn)}
\begin{document}
\setkeys{Gin}{draft=false}

\title{Magnetohydrodynamic Simulation of the Interaction between Interplanetary Strong Shock and Magnetic
Cloud and its Consequent Geoeffectiveness}

\author{Ming Xiong, Huinan Zheng, Yuming Wang, and Shui Wang}
\affil{CAS Key Laboratory for Basic Plasma Physics, School of Earth and Space Sciences, University of
Science and Technology of China, Hefei, Anhui 230026, China}



\begin{abstract}
Numerical studies have been performed to interpret the observed
``shock overtaking magnetic cloud (MC)'' event by a 2.5
dimensional magnetohydrodynamic (MHD) model in heliospheric
meridional plane. Results of an individual MC simulation show that
the MC travels with a constant bulk flow speed. The MC is injected
with very strong inherent magnetic field over that in the ambient
flow and expands rapidly in size initially. Consequently, the
diameter of MC increases in an asymptotic speed while its angular
width contracts gradually. Meanwhile, simulations of MC-shock
interaction are also presented, in which both a typical MC and a
strong fast shock emerge from the inner boundary and propagate
along heliospheric equator, separated by an appropriate interval.
The results show that the shock firstly catches up with the
preceding MC, then penetrates through the MC, and finally merges
with the MC-driven shock into a stronger compound shock. The
morphologies of shock front in interplanetary space and MC body
behave as a central concave and a smooth arc respectively. The
compression and rotation of magnetic field serve as an efficient
mechanism to cause a large geomagnetic storm. The MC is highly
compressed by the the overtaking shock. Contrarily, the transport
time of incidental shock influenced by the MC depends on the
interval between their commencements. Maximum geoeffectiveness
results from that when the shock enters the core of preceding MC,
which is also substantiated to some extent by a corresponding
simplified analytic model. Quantified by $Dst$ index, the specific
result gives that the geoeffectiveness of an individual MC is
largely enhanced with 80\% increment in maximum by an incidental
shock.
\end{abstract}
\begin{article}


\section{Introduction}
Coronal mass ejection (CME) is one of the most frequently eruptive
phenomena in solar atmosphere, which causes significant changes in
coronal structure accompanied by observable mass outflow. A great
deal of CME observation data has been accumulated by spacecraft
OSO-7, Skylab, P78-1, SMM, ISEE3, Helios, Yohkoh, SOHO, Ulysses,
Wind, ACE et al. over the past 30 years. A typical CME is launched
into interplanetary (IP) space with magnetic flux of
$10^{23}$maxwell and plasma mass of $10^{16}$g
\citep{Gosling1990,Webb1994}. The ``solar flare myth'' that CMEs
have no fundamental association (in terms of cause and effect)
with flares \citep{Gosling1993,Gosling1995} is quite controversial
\citep[e.g.,][]{Svestka1995,Dryer1996}. It is more favorable of
the equal importance of CME and flare concerning the source of IP
transient disturbances and non-recurrent geomagnetic storms
\citep{Dryer1996}. Statistical research shows that nearly half of
all CMEs form magnetic clouds (MCs) in IP space
\citep{Klein1982,Cane1997}. MC is very concerned in space
community, because its regular magnetic field with large southward
magnetic component always leads to geomagnetic storm. The
characteristics of MCs, as defined by \citet{Burlaga1981}, are
enhanced magnetic field, smooth rotation of the magnetic field,
low proton temperature, and a low ratio of proton thermal to
magnetic pressure $\beta_p$. Many studies modeled an MC by an
ideal local cylinder with a force-free field
\citep[e.g.,][]{Goldstein1983,Burlaga1988,Farrugia1993,Kumar1996,Osherovich1997},
though in real situation an MC should probably be a curved
loop-like structure with its feet connecting to the solar surface
\citep{Larson1997}. Numerical simulations have been carried out to
investigate the behavior of isolated loop-like MCs with various
magnetic field strengths, axis orientations and speeds, based on
the flux rope model
\citep[e.g.,][]{Vandas1995,Vandas1996a,Vandas1996b,Vandas1996c,Vandas1997a,Vandas1997b,Vandas2000,
Vandas2002,Groth2000,Odstrcil2002,Schmidt2003,Vandas2003,Manchester2004a,Manchester2004b}.
A great consistency was found between the in-situ observations,
theoretical analyses and numerical simulations.

Recent studies have focused on the existence of more complex structure, with less defined characteristics
and a possible association with interactions among CMEs, shocks, MCs, and corotating regions, such as
complex ejecta \citep{Burlaga2002}, multiple MCs \citep{Wang2002,Wang2003a}, shock-penetrated MCs
\citep{Wang2003b,Berdichevsky2005}, and so on. Most of the different physical phenomena, which are likely
to occur during the propagation of a following faster CME overtaking a preceding slower CME, have been
studied by both 2.5-dimensional (2.5D) and 3-dimensional (3D) magnetohydrodynamic (MHD) numerical
simulations: the interaction of a shock wave with an MC \citep{Vandas1997a,Odstrcil2003}, the interaction
of two MCs \citep{Odstrcil2003,Gonzalez-Esparza2004,Lugaz2005,Wang2005}, and the acceleration of electrons
associated with the shock-cloud interaction \citep{Vandas2004}.

The establishment of space weather forecasting system is ongoing
as urgently needed by human civilization. Numerical MHD model may
play a critical role in it \citep{Dryer1998}. IP medium is a
pivotal node in cause-effect chains of solar-terrestrial
transporting events. The correlation between $Dst$ index and
various IP parameters have been comprehensively studied
\citep[e.g.,][]{Burton1975,Vassiliadis1999} and applied in related
numerical simulations \citep[e.g.,][]{Vandas2003}. Moreover, some
observation-data-driven numerical models have already been applied
in the real time ``fearless forecasting": (1) HAF
(Hakamada-Akasofu-Fry) model based on kinetics
\citep{Fry2001,Fry2005,Intriligator2005,McKenna-Lawlor2005}; (2)
STOA (Shock Time of Arrival) based on classical self-similarity
blast wave theory \citep{Smart1985}; (3) ISPM (Interplanetary
Shock Propagation Model) based on 2.5D MHD simulation
\citep{Smith1990}; (4) an ensemble of above three models
\citep{Dryer2001,Dryer2004,McKenna-Lawlor2002,Fry2003,Fry2004}.

The observed ``shock overtaking MC'' event complicates IP
dynamics. With an enough strong magnitude, a fast shock can
propagate through a low $\beta$ MC and survive as a discontinuity
in front part of the MC. It can even penetrate the MC and merge
with the original MC-driven shock into a stronger compound shock.
The evolution stages of MC-shock interaction detected by Wind and
ACE spacecraft at 1 AU may be reduced into two categories: (1)
shock still in MC, such as October 3-6 2000 and November 5-7 2001
events \citep{Wang2003b}; (2) shock ahead of MC after completely
penetrating it, such as March 20-21 2003 event
\citep{Berdichevsky2005}. Ruling out the possibility of weak shock
dissipation in low $\beta$ MC plasma, the MC-shock compound at 1
AU changes from category 1 to 2, as their eruption interval
decreases at solar corona. MC-shock interaction is also an IP
cause of large geomagnetic storms \citep{Wang2003b,Wang2003c}.
Obviously MC with a penetrating shock at various stages may result
in different geoeffectiveness.

In this paper, studies are presented to understand the dynamic
process of the ``shock overtaking MC'' event and its effect on
geomagnetic storm strength by numerical simulation based on a 2.5D
ideal MHD model. A brief description of the MHD equations and the
numerical scheme used to solve them, as well as the steady state
solar wind, the MC configuration and shock specification, is given
in Section~\ref{Sec:Method}. Simulation results of an individual
MC are described in Section~\ref{Sec:MCOnly}. Results of MC-shock
interaction are discussed and analyzed in
Section~\ref{Sec:MCShock}. The geoeffectiveness of MC-shock
interaction is discussed in Section~\ref{Sec:Geoeffect}. Finally,
conclusions are summarized in Section~\ref{Sec:Conclusion}.

%
%
\section{Numerical MHD Model}\label{Sec:Method}

\subsection{Governing MHD Equations}

The macro-scope behavior of magnetized plasma can be well described with MHD equations by using the
conservation laws, supplemented by the equation of state of fluids and divergence-free condition of
magnetic field. Since IP magnetic field (IMF) co-rotates with the Sun, it is convenient to adopt a
co-rotating coordinate system, in which the fluid velocity is parallel to the magnetic field. With the
assumption of an ideal gas with a polytropic index $\gamma = 5/3$ and neglecting the effects of viscosity,
electrical resistivity, and thermal conduction, the ideal MHD equations are written as follows (cf.
\citet{Jeffrey1964}).
\begin{eqnarray}
\label{Eqn:Mass} && \frac{\partial \rho}{\partial t} + \nabla \cdot (\rho
   \mathbf{v})
      = 0 \\
&& \frac{\partial (\rho \mathbf{v})}{\partial t} + \nabla
    \cdot \left[\rho \mathbf{v} \mathbf{v} +  \left(p+ \frac{1}{8\pi}B^2 \right) I - \frac{1}{4\pi}\mathbf{B}\mathbf{B} \right]= \mathbf{f}\\
&& \frac{\partial \mathbf{B}}{\partial t} - \nabla \times (\mathbf{v}
     \times \mathbf{B})
      = 0 \\
&& \frac{\partial W}{\partial t} + \nabla \cdot \left[
   \left( \frac{\gamma}{\gamma-1}p + \frac{1}{2}\rho v^2
   \right)\mathbf{v} + \frac{1}{4\pi} \mathbf{B} \times (\mathbf{v} \times \mathbf{B}) \right
   ] \nonumber\\
\label{Eqn:Energy} && \qquad = \mathbf{f} \cdot \mathbf{v}
\end{eqnarray}
with
\begin{eqnarray}
 && \mathbf{f}= - \rho \left[\frac{g R_s^2}{r^2} \frac{\mathbf{r}}{r} + \mathbf{\Omega} \times
(\mathbf{\Omega} \times \mathbf{r}) + 2(\mathbf{\Omega} \times \mathbf{v})   \right]
 \nonumber\\
 && W = \frac{1}{2} \rho v^2 + \frac{1}{8\pi} B^2 +
                  \frac{p}{\gamma-1}.  \nonumber
\end{eqnarray}
Where $\rho$ is the plasma mass density, $\mathbf{v}$ the plasma
velocity, $\mathbf{B}$ the magnetic field, $p$ the plasma pressure
(sum of electron and proton pressures), $\mathbf{\Omega}$ the
angular speed of solar rotation ($=2.9 \times 10^{-6}\mbox{
rad/s}$), $I$ the unit matrix, $R_s$ the solar radius, $g$ the
gravitational acceleration at the solar surface.
Equations~(\ref{Eqn:Mass})-(\ref{Eqn:Energy}) are expressed in
spherical coordinate system $(r, \theta, \varphi)$, dealing with
2.5D problems in the meridional plane. Namely, the partial
derivatives of all dependent variables with respect to azimuthal
angle $\varphi$ are zero.

\subsection{Computational Techniques}\label{Sec:ComTec}

The mathematical connotation of shock overtaking MC belongs to
high resolution problems for the interaction between discontinuity
and complex smooth structure. Total variation diminishing (TVD)
scheme, a shock-capturing method, is applied to numerically solve
MHD equations \citep{Harten1983, Ryu1995}, which possesses a
formal accuracy of the second order in smooth flow regions except
at extreme points. An 8 wave model \citep{Powell1995} is adopted
to guarantee divergence-free condition of magnetic field.

Furthermore, the magnetic flux function $\psi$ is introduced to
ensure the accuracy of magnetic field in the region near the shock
front and the MC, which satisfies
\begin{eqnarray}
 && \frac{\partial \psi}{\partial t} + v_r \frac{\partial \psi}{\partial
    r}+\frac{v_\theta}{r} \frac{\partial \psi}{\partial\theta}=0
    \label{Eqn:psi}
\end{eqnarray}
with
\begin{eqnarray}
 && \mathbf{B}= \left(\frac{1}{r^2 \sin\theta} \frac{\partial \psi}{\partial\theta}, -\frac{1}{r \sin\theta} \frac{\partial\psi}{\partial r},
    B_\varphi \right)  \label{Eqn:psiB}.
\end{eqnarray}
Equation~(\ref{Eqn:psi}) is solved by fifth order weighted
essentially non-oscillation (WENO) scheme \citep{Shu1997} and the
meridional components of magnetic field are updated by $\psi$ in
equation~(\ref{Eqn:psiB}). In addition, special techniques in the
numerical simulations of magnetic flux rope
\citep{Hu2003,Zhang2005} are introduced here, which eliminate the
numerical reconnection across the heliosphere current sheet (HCS)
and guarantee the conservations of mass, axial and toroidal
magnetic fluxes of magnetic rope.

For simulations in this paper, computational domain is taken to be $25 R_s \le r \le 300 R_s$, $0^\circ \le
\theta \le 180^\circ$ and discretized in meshes evenly spaced with $\Delta r=1.5 R_s$  and $\Delta
\theta=1.5^\circ$. To avoid the complex boundary conditions associated with transonic flow, the inner
boundary of computational domain is chosen so that the solar wind speed has already exceeded the fast
magnetoacoustic speed. Since all waves are entering the domain at the inner boundary $(r=25R_s)$, all
quantities can be specified independently. While linear extrapolations are exerted at the outer boundary
$(r=300R_s)$ where all waves exit the domain. Symmetric conditions are used at latitudinal directions.

\subsection{Ambient Solar Wind Equilibrium}\label{Sec:AmbSWE}

Ambient solar wind equilibrium is obtained simply by specifying
the inner boundary conditions. A unique steady state solar wind
solution is obtained after $\sim 120$ hours by fixing a set of
parameters at the inner boundary, with proton number density
$N_p=550~\mbox{ cm}^{-3}$, radial solar wind speed $v_r=375 \mbox{
kms}^{-1}$, magnetic field strength $B=400 \mbox{ nT}$, the plasma
beta (defined as the ratio of plasma thermal to magnetic pressure)
$\beta=\frac{8\pi p}{B^2} = 0.23$, as well as the conditions
$B_\theta = 0$ and $\mathbf{v} \parallel \mathbf{B}$. The
configuration is quite similar to that by \citet{Wang2005}, with
its typical values at $25R_s$ (the inner boundary) and $213R_S$
(near the earth orbit) listed in Table~\ref{Tab1}. An HCS is
introduced by simply reversing the magnetic field across the
equator, i.e. magnetic field directs outwards (inwards) in
southern (northern) semi-heliosphere. Theoretically, the HCS is an
ideal tangential discontinuity in MHD macro-scale, but it is here
smeared out over several grids by numerical diffusion. However
this slightly smeared structure is quite similar to the
configuration that an HCS is embedded in a relatively thicker
heliospheric plasma sheet (HPS), which is substantiated by space
observation during solar minimum \citep{Winterhalter1994}. In
addition, the equilibrium here does not resemble the bimodal
nature of the solar wind with fast flow over the poles and slow
flow at low latitudes. We argue that this will not distort the
fundamental physical process of the MC-shock interaction, which
locates mainly at low latitudes. The ambient equilibrium is
described as slow solar wind astride HCS--HPS.

\subsection{Specification of MC and Shock Emergences}

Specific methods for MC injection by \citet{Vandas1995} and fast
shock injection by \citet{Hu1998,Hu2001} are applied in our
simulation through the inner boundary condition modification. Once
MC or shock is completely emerged into IP medium, the original
inner boundary condition as mentioned in Section~\ref{Sec:AmbSWE}
is restored.

The magnetic field configuration of an MC is described as Lundquist solution in local cylindrical
coordinate $(R, \Phi, Z)$ \citep{Lundquist1950}.
\begin{eqnarray}
   \left\{ \begin{array}{ll}
        B_R=0   \\
        B_\Phi = B_0 H J_1(\alpha R) \\  \label{Eqn:Lundquist}
        B_Z = B_0 J_0(\alpha R)  \\
           \end{array} \right.
\end{eqnarray}
where $B_0$ specifies the magnetic field magnitude at MC core, $H$ is the magnetic helicity,
$\alpha=2.4/R_m$ and $R_m$ is MC radius. With given emergence time $t_m$, mass $M_m$, speed $v_m$, radius
$R_m$, plasma $\beta$, and helicity $H$ together with above magnetic configuration, an MC is uniquely
determined. It is unrealistic to approximate the 3D structure of an MC that is rooted deeply in solar
surface in 2.5D coordinate system. However, regarding MC as a section of the 3D magnetic loop, its dynamic
characteristics could still be reflected by a 2.5D numerical simulation.

An incidental fast shock is characterized by several parameters:
its emergence time $t_{s0}$, the latitude of its center
$\theta_{sc}$, the latitudinal width of its flank $\Delta
\theta_s$, the maximum ratio of total pressure (sum of thermal and
magnetic pressures) at shock center $R^*$, the duration of growth,
maintenance and recovery phases ($t_{s1}$, $t_{s2}$, $t_{s3}$).
The ratio of total pressure decreases from $R^*$ at center to $1$
at both flank edges as cosine function of the angle. It varies
linearly with time during the growth and recovery phases of shock
disturbance. Given the upstream state at the inner boundary and
$R^*$, downstream state is derived by Rankine--Hugoniot
relationship. The introduced shocks in our simulation are strong
enough to be faster than the local magnetosonic speed at all time.
A shock can be formed closer to the sun, below the
usually-computed steady-state critical points. Many solar
observations show that shock can be formed below Alfv\'en critical
point which is below the inner boundary of the computational
domain \citep[e.g.,][]{Cliver2004,Raouafi2004,Cho2005}.

\section{Propagation of an Individual MC (Case A)}\label{Sec:MCOnly}

We present here an individual MC simulation firstly, to manifest
its characteristics, as well as for comparison with MC-shock
interaction in the next section. The MC emerges along HCS from the
inner boundary. It takes the following parameters referring to
Equation (\ref{Eqn:Lundquist}),
\begin{eqnarray}
    && R_m = 5 R_s, \quad B_0 = 1700 \mbox{ nT}, \quad H=1 \nonumber
\end{eqnarray}
and
\begin{eqnarray}
    && v_m = 530 \mbox{ kms}^{-1}, \quad M_m = 4.8 \times 10^{12} \mbox{ kg}, \quad \beta = 0.02. \nonumber
\end{eqnarray}
Magnetic flux function $\psi$ (cf. Equation~(\ref{Eqn:psiB})) is
$1.51 \times 10^{14} \mbox{ Wb}$ at the core of MC, comparing with
$1.12 \times 10^{14}$ and $0~\mbox{ Wb}$ in HCS and heliospheric
poles, respectively. The axial magnetic flux of MC is calculated
to be $2.5 \times 10^{13} \mbox{ Wb}$. This MC and its surrounding
IMF have the same magnetic polarity in meridional plane.

\begin{figure}
\noindent
    \includegraphics{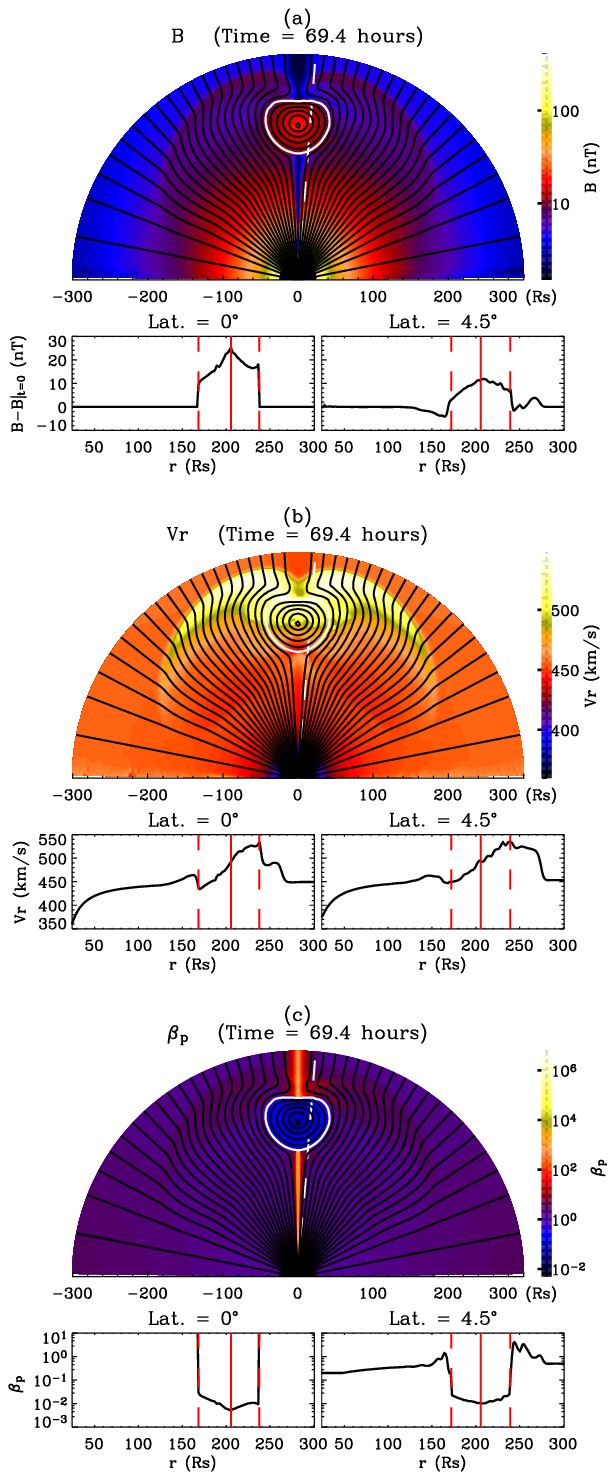}
\caption{} \label{case-A}
\end{figure}

The simulation of an MC passing nearby the Lagrangian point (L1)
is shown in Figure~\ref{case-A}. Under each image are two
corresponding radial profiles by cutting right through $0^\circ$
(noted by $~\mbox{Lat.}=0^\circ$) and $4.5^\circ$ (white dashed
lines in the images, noted by $~\mbox{Lat.}=4.5^\circ$) away from
the equator. The magnitude of magnetic field in radial profile is
given by subtracting its corresponding initial value of ambient
equilibrium. The body of MC is identified to be enclosed by a
white solid line in the images and between two dotted lines in
attached profiles. The white solid line is determined by the
magnetic flux function ($\psi$) value in the equator plus a small
increment. This line lies right inside the MC boundary which has
the flux function value equal to that in the equator. The MC core
is determined by the maximum value of $\psi$. Magnetic field
configuration is superimposed upon the images. As shown in the
Figure~\ref{case-A}, the MC ejection into ambient solar wind
results in two distinct interaction regions: (1) an MC envelope
composed of IMF draping around self-enclosed MC surface; (2) a
shock front and its associated sheath ahead of MC body formed by
the compression of the high-speed MC. A concave is formed at the
MC-driven shock front across the HPS, as clearly seen in
Figure~\ref{case-A}(b), which is also substantiated by IPS
(interplanetary scintillation) observation \citep{Watanabe1989}
and shock-related simulations
\citep{Odstrcil1996a,Odstrcil1996b,Hu2001}. The characteristics of
shock front are caused by the particular HCS-HPS structure in
heliosphere. Shock degenerates abruptly into hydrodynamic shock
due to nearly vanishing magnetic field at neutral current sheet.
So the fastest and strongest shock front locates on the edges of
HPS instead of being right in HCS. The angular width of shock
front is much larger than that of its driver--MC body. Monotonic
decrease of bulk flow speed $v_r$ in the MC, as seen from
Figure~\ref{case-A}(b), implies continuous MC expansion through IP
space. Moreover, many other MC characteristics are also manifested
in agreement with the observations. These characteristics maintain
till the MC propagates beyond the outer boundary.

\begin{figure}
\noindent
    \includegraphics[width=20pc]{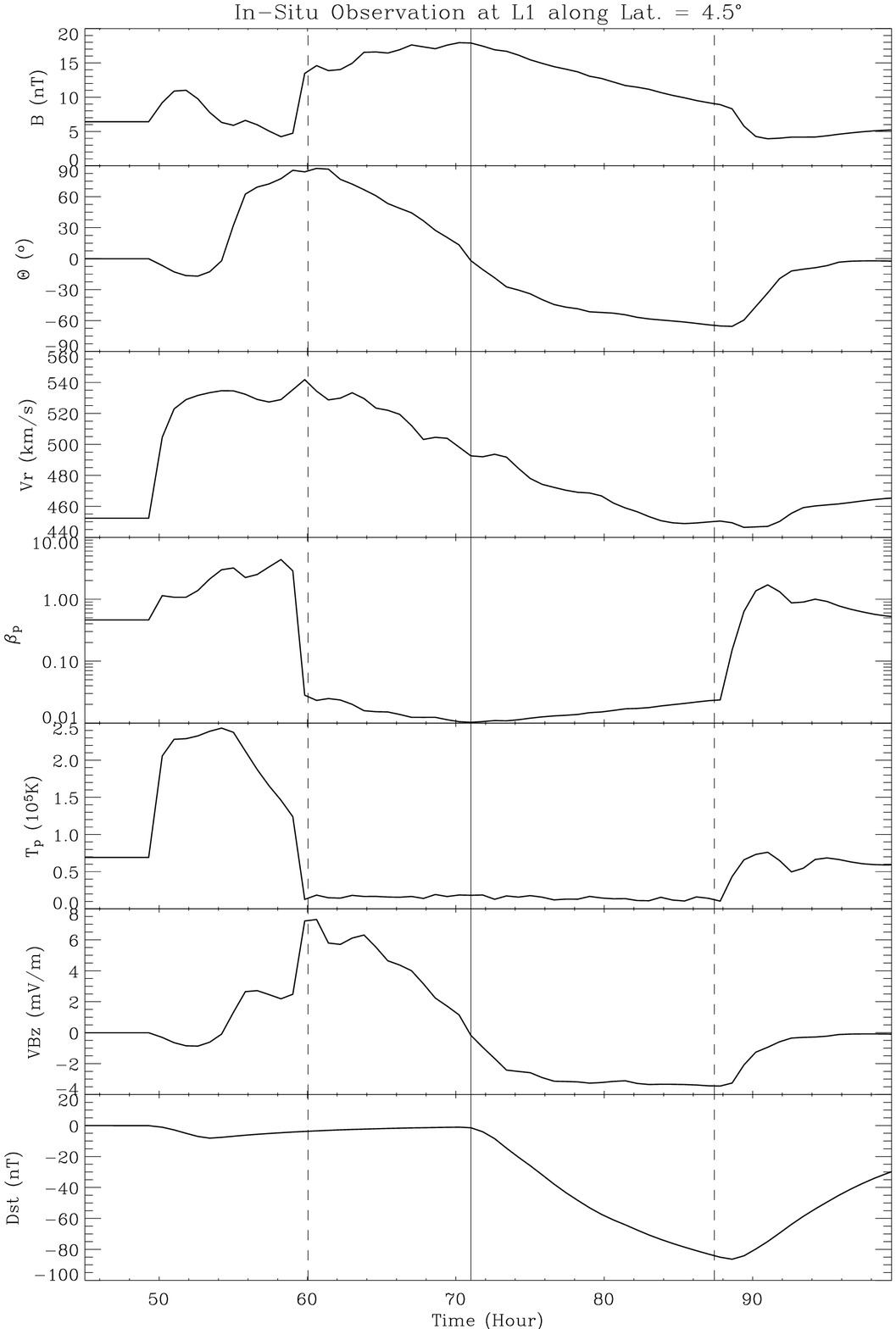}
\caption{}\label{mc.L1}
\end{figure}

The in-situ measurement along $~\mbox{Lat.}=4.5^\circ$ by a
hypothetic spacecraft at L1 is shown in Figure~\ref{mc.L1}.
Typical MC characteristics, such as enhanced magnetic magnitude
$B$ (also in attached profiles in Figure \ref{case-A}(a)), smooth
rotation of magnetic field $\Theta$, a low concave of proton
temperature $T_p$ and proton beta $\beta_p$ (also in attached
profiles in Figure \ref{case-A}(c)), continuous decrease of bulk
flow speed $v_r$ (also in attached profiles in Figure
\ref{case-A}(b)) and so on, are reproduced. A sheath ahead of the
MC with high temperature and high speed is clearly seen too. The
shock front, the leading, central and trailing parts of MC pass by
L1 at 49.3, 60, 71 and 87.4 hours successively. The MC event at L1
lasts 27.4 hours, with a maximum magnetic field magnitude ($17.9
~\mbox{nT}$) and a minimum southward component ($-7.7 ~\mbox{
nT}$). The geomagnetic effect of the simulated MC event is
evaluated by $Dst$ index, as applied by \citet{Wang2003c} using
formula $\frac{d Dst(t)}{dt}=Q(t)-\frac{Dst(t)}{\tau}$
\citep{Burton1975}, where the coupling function $Q=VB_s$ (here $V$
is evaluated with $v_r$, $B_s=\min(B_z,0)$ and $B_z$ is the $z$
component of magnetic field) and the diffusion time scale $\tau=8
\;\mbox{hours}$. The MC center approaches L1 71 hours after its
departure from the inner boundary, and the value of $Dst$ index
decreases monotonically to its minimum $-86\;\mbox{nT}$ shortly
afterwards (at 88.6 hours). In addition, the draping IMF within
MC-driven sheath is mainly northward. This is why the compressed
magnetic field in the sheath does not cause significant $Dst$
disturbance in our simulation.

\section{Interaction between a Fast Shock and a Preceding MC}\label{Sec:MCShock}

\subsection{Case B}

Shock compression is an efficient mechanism for the enhancement of
southward component of magnetic field and, hence, serves as an IP
cause of large geomagnetic storms \citep{Wang2003b, Wang2003c}.
The subsequent numerical simulations aim to quantify
geoeffectiveness of a shock overtaking an MC in detail.

To investigate the interaction between a fast forward shock and a
preceding MC, a shock centered at HCS ($\theta_{sc}=0^\circ$) is
introduced from the inner boundary to pursue the previous occurred
MC. The MC in this case is identical with that in Case A. The
shock emerges at $t_{s0} = 41\mbox{ hours}$ with its center on the
equator, and other parameters
\begin{eqnarray}
    \Delta\theta_s = 6^\circ, \quad R^* = 24, \quad t_{s1} = 0.3 \mbox{ hr}, \quad t_{s2} = 1 \mbox{ hr},
    \quad t_{s3} = 0.3 \mbox{ hr}. \nonumber
\end{eqnarray}
One can find that the maximum shock speed is $1630 \mbox{
kms}^{-1}$ from the above quantities by the shock relation. The
ratio of total pressure decreases from $R^*$ at the equator to 1
at $\pm 6^\circ$ aside via a cosine function. The temporal extent,
as already specified, can be described as being trapezoidal as
done, for example, by \citet{Smith1990} in the ecliptic plane.

\begin{figure}
\noindent
   \includegraphics[height=39pc,angle=90]{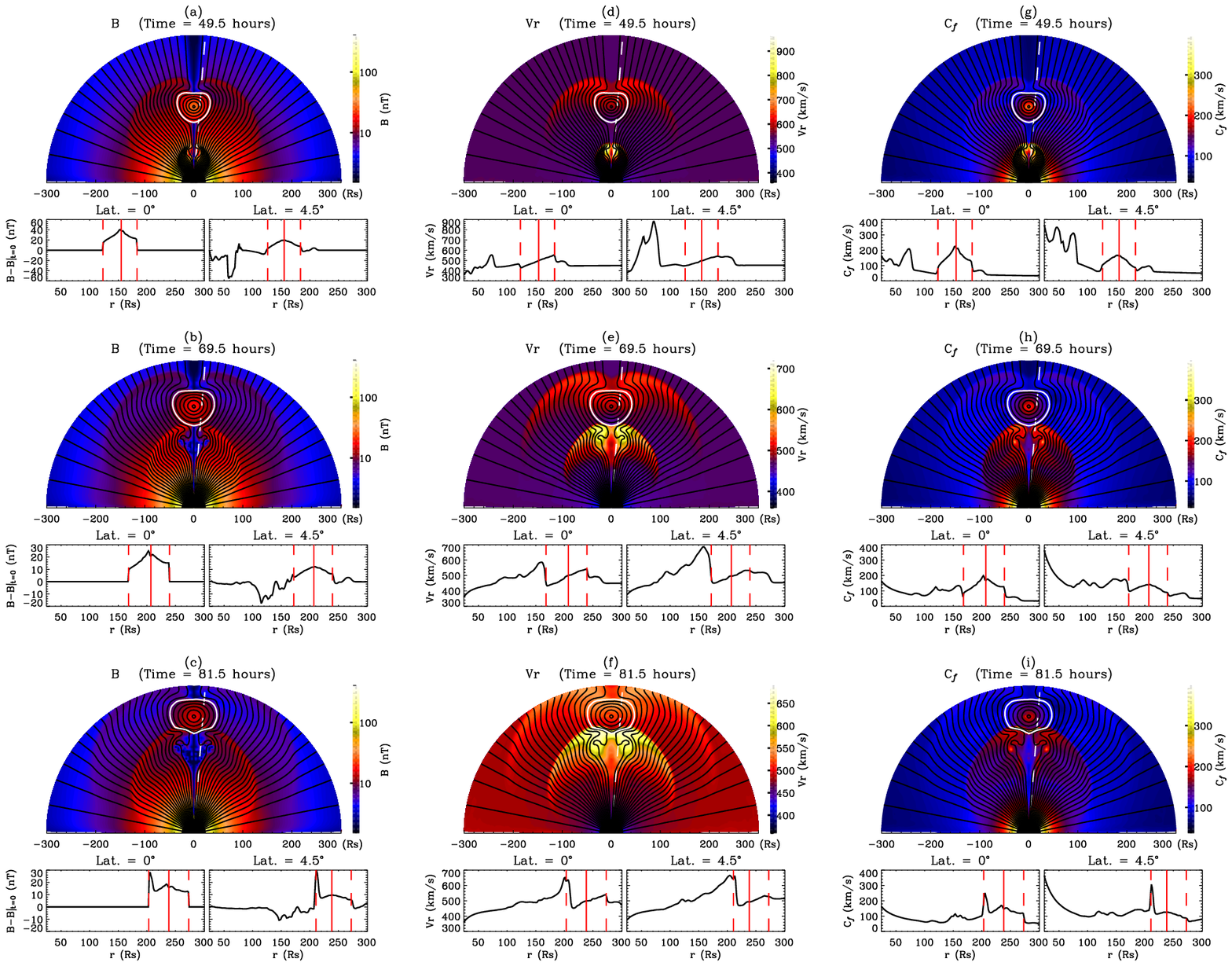}
\caption{} \label{case-B}
\end{figure}

The detail process of MC-shock event is elucidated in
Figure~\ref{case-B}. The incidental shock aphelion and the MC core
arrive at $80R_s$ and $155R_s$ respectively in 49.5 hours, shown
in Figure~\ref{case-B}(a), (d) and (g). The morphology of shock
front has a dimple across HCS, similar to that of MC-driven shock
mentioned previously. In the downstream of shock front, the flow
speed reaches its maximum value, $900 \mbox{ kms}^{-1}$,
$4.5^\circ$ away from HCS, much greater than that right at HCS,
which is $560 \mbox{ kms}^{-1}$. Comparing with the preceding MC,
which has a peak speed of $540 \mbox{ kms}^{-1}$ only, the
overwhelming forward shock will soon collide with MC body.
Moreover, the tangential magnetic field component increases as
fast shock passes by. So that IMF in either semi-heliosphere is
deflected to the pole by the impaction of shock propagation. As a
result, a ``magnetic vacuum" with weaker magnetic field strength
nearby HCS is formed just behind the shock front, as indicated in
Figure~\ref{case-B}(a)-(c). The shock just catches up with the
inner boundary of MC at 69.5 hours (Figure~\ref{case-B}(b), (e)
and (h)). In addition, as shown in Figure~\ref{case-B}(h), the
radial characteristic speed of fast mode wave $c_f$ is very large
in the MC body with low $\beta$. It increases steadily from $100
\mbox{ kms}^{-1}$ at MC boundary to $200 \mbox{ kms}^{-1}$ in
maximum at MC core. There is also a peak value $180 \mbox{
kms}^{-1}$ for $c_f$ within the shock sheath. Meanwhile, $v_r$ in
the MC decreases monotonically from $540 \mbox{ kms}^{-1}$ to $430
\mbox{ kms}^{-1}$ along Lat. $=0^\circ$, as seen in
Figure~\ref{case-B}(e). MC-shock collision is pregnant at this
critical time. Moreover (1) the aphelion of shock front locates on
the edges of HPS instead of being right at HCS due to its concave
morphology; (2) the incidental shock in HCS is virtually a
relatively weaker hydrodynamic shock. Though the center of shock
front is along HCS, the most violent collision can be witnessed
consequently at shock aphelion rather than in HCS when the shock
and MC collide with each other. A sharp discontinuity has already
been formed in the rear part of MC at 81.5 hours
(Figure~\ref{case-B}(c), (f) and (i)). The compression along HCS
(Lat. $=0^\circ$) is less significant than that along Lat.
$=4.5^\circ$. The influence of fast shock upon MC could be reduced
into two aspects: (1) enhancement of magnetic field magnitude; (2)
rotation of magnetic field. As shown in Figure~\ref{case-B}(c),
the maximum value of magnetic field enhancement $(B-B|_{t=0})$ is
30 nT in compressed region, much larger than that at MC core, 18
nT. Compressed magnetic field lines are very flat and point nearly
southwards. Both effects result in a minimum southward magnetic
field $B_z$ with -33 nT at MC tail. Furthermore, $c_f$ is enhanced
simultaneous during shock compression, as seen from
Figure~\ref{case-B}(i). In contrast with $160 \mbox{ kms}^{-1}$ at
MC core, $c_f$ at MC tail has jumped to $300 \mbox{ kms}^{-1}$.
However, the strong shock is not counteracted completely by the
enhanced $c_f$ in the MC medium. Its propagation in MC would not
be stopped or diffused despite MC resistance. In addition, the
domain of so-called ``magnetic vacuum" behind shock front is
magnified during the process of shock overtaking rear part of MC,
because the MC, an enclosed magnetic loop, serves as an obstacle
in front of shock. The MC just passes by L1 at 82 hours. Though
shock continues to penetrate MC into a deeper position, MC-shock
compound structure will no longer cause the geoeffectiveness
shortly after it passes by the orbit of earth.

\begin{figure}
\noindent
   \includegraphics[width=20pc]{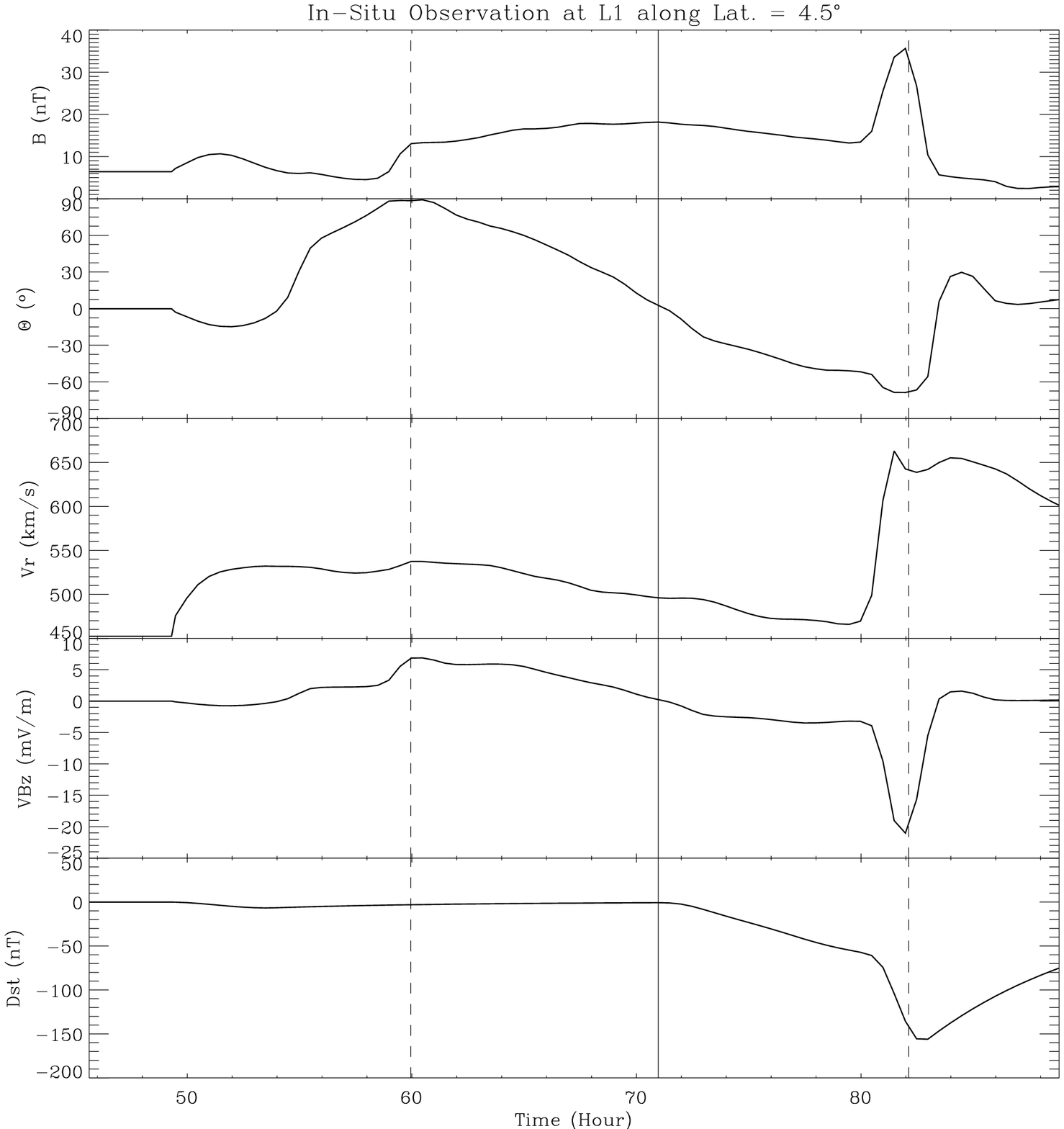}
\caption{} \label{mc-shock-B.L1}
\end{figure}

Similar to that in Case A, the simulated data at L1 in time
sequence are shown in Figure~\ref{mc-shock-B.L1}. A bump on the
tail of MC is obviously found around 81 hours, with a peak speed
$660 \mbox{ kms}^{-1}$ larger than $540 \mbox{ kms}^{-1}$ at the
head of MC. As a consequence, $VB_z$ jumps from $-4 \mbox{
mVm}^{-1}$ to $-21 \mbox{ mVm}^{-1}$ for less than 2 hours. By
comparing with Figure~\ref{mc.L1}, one can see from
Figure~\ref{mc-shock-B.L1} that the index of geomagnetic storm
$Dst$ is -156 nT in MC-shock compound structure, much greater than
-86 nT in the corresponding individual MC event. Moreover, the
rear boundary of MC leaves L1 5.3 hours earlier than that in Case
A. MC is highly compressed in its rear part by the shock.

\subsection{Case C}

To further explore the features of MC-shock interaction in
solar-terrestrial range, we give another case of simulation (Case
C) where the shock ultimately penetrates the preceding MC nearby
L1. It is straightforward to schedule an earlier shock emergence.
The shock emergence time $t_{s0}$ is modified to be 10 hours
compared with 41 hours in Case B. All other parameters are the
same as those in Case B.

\begin{figure}
\noindent
    \includegraphics[width=20pc]{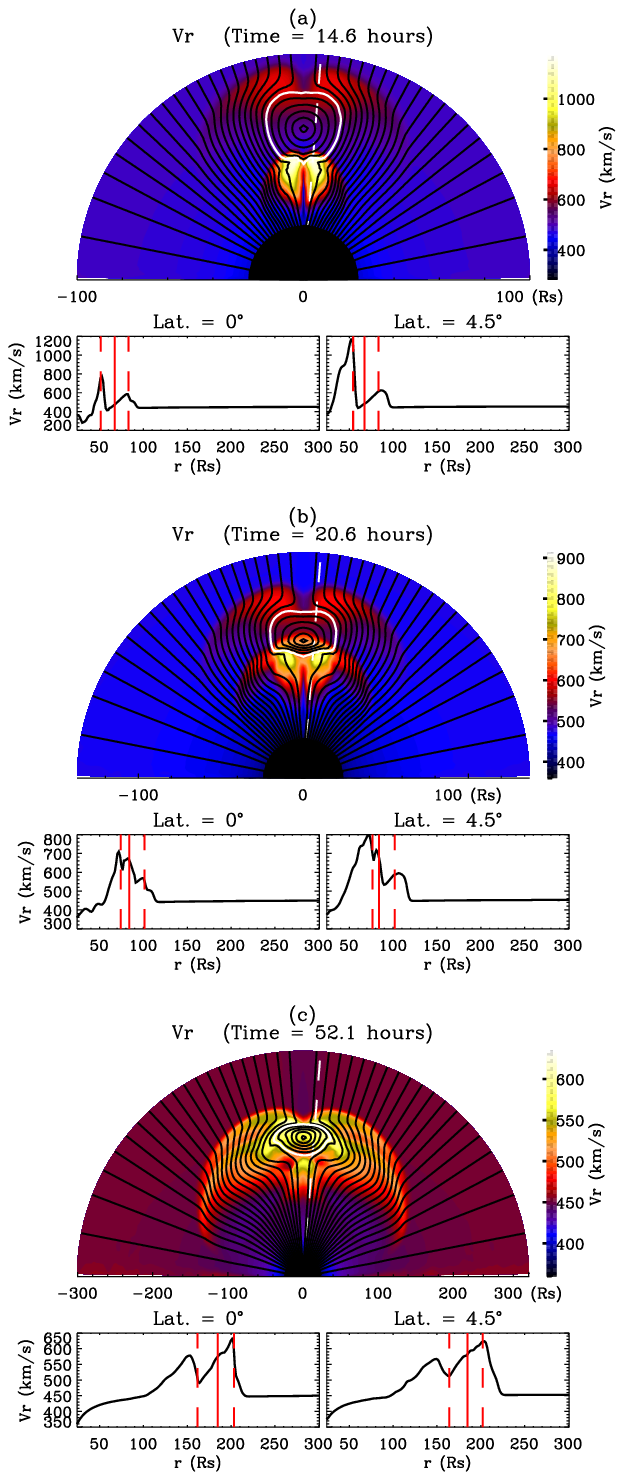}
\caption{} \label{Fig:CaseC}
\end{figure}

Only the evolution of $v_r$ is given in Figure~\ref{Fig:CaseC}, to
visualize the concerned MC-shock complex structure. Once the fast
shock advances deeply into the MC, the latter, superseding the
ambient IP space, serves as a medium for the shock propagation.
Since HCS does not exist in the MC, what ensues is the
disappearance of the HCS-associated concave. The morphology of
shock front is a smooth arc in the highly compressed rear part of
MC at 20.6 hours (Figure~\ref{Fig:CaseC}(b)). When the shock
penetrates and emerges from the MC, HCS-HPS structure re-plays an
important role in shock propagation. The smooth arc quickly turns
into a concave across the equator with respect to shock front at
52.1 hours, as indicated clearly by Figure~\ref{Fig:CaseC}(c).
This newly emerged fast shock from the MC will gradually merges
with the preceding MC-driven shock into a stronger fast shock by
nonlinear interaction. Moreover, sheath width, defined by the
radial distance along the equator between MC-driven shock front
and the outer MC boundary, is $10R_s$ in Case C, only half of that
in Case A, $20R_s$. Compared with Case A, several distinct
differences are easily discriminated in Case C to emphasize the
shock impact: (1) the geometry of MC boundary changes in the shape
from quasi-circle to oblate ellipse; (2) MC is highly compressed;
(3) the width of MC-driven sheath is significantly narrowed.

\begin{figure}
   \includegraphics[width=20pc]{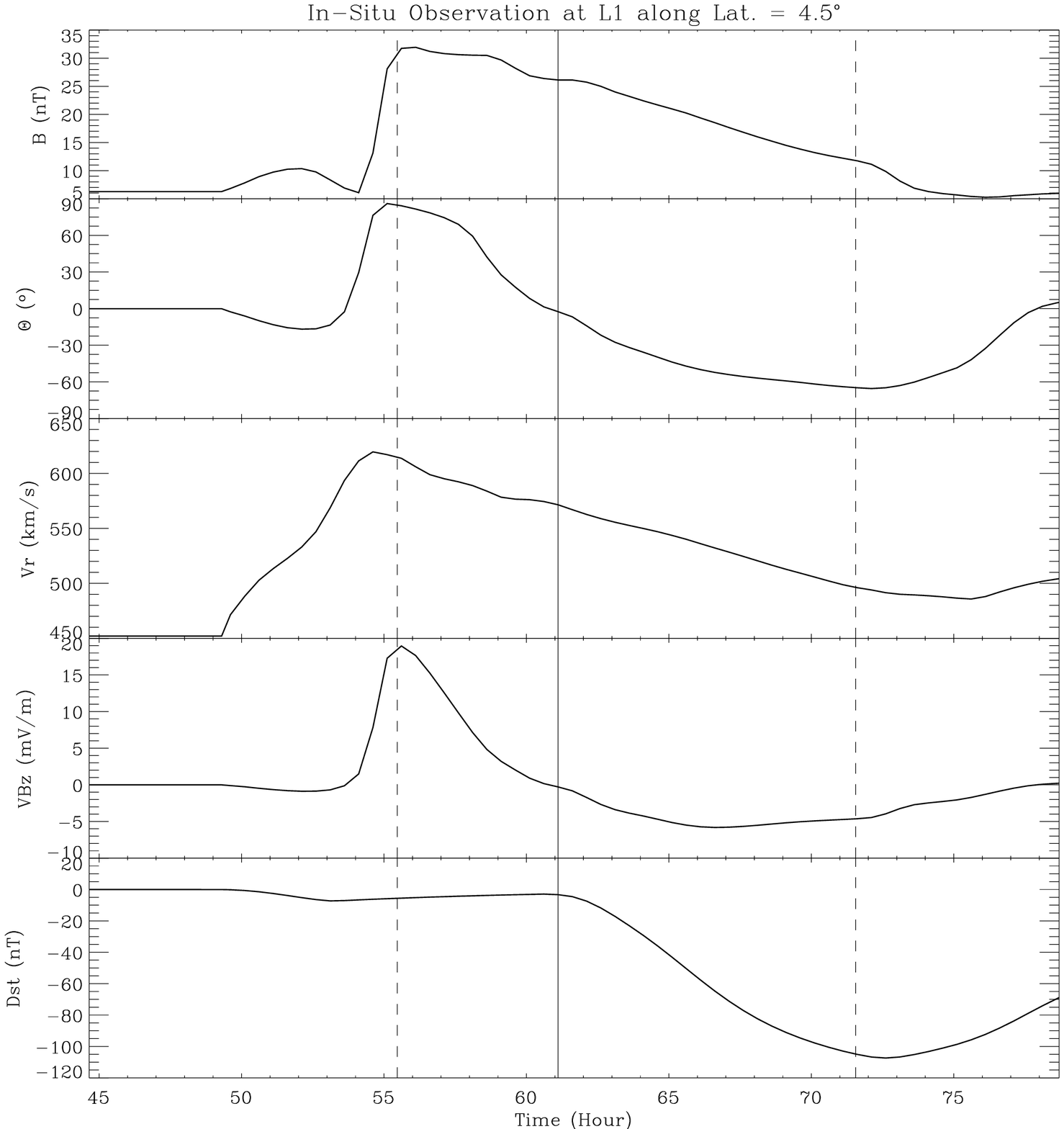}
\caption{} \label{mc-shock-C.L1}
\end{figure}

The hypothetic in-situ measurement at L1 along Lat. $=4.5^\circ$
is plotted in Figure~\ref{mc-shock-C.L1} in contrast. The outer
boundary, the center, and the inner boundary of MC arrive at L1 at
55.5, 61 and 71.5 hours successively, which are 4.5, 10 and 15.9
hours earlier than those in Case A indicated by
Figure~\ref{mc.L1}. In presence of the shock penetration, the
duration of MC passing across L1 is shortened by 11.4 hours. No
extremum of speed profile is found inside the MC because the shock
has moved out of it. Judged only from single-spacecraft in-situ
observation, as seen from Figure~\ref{mc-shock-C.L1}, it resembles
quite likely an individual MC event with a peak speed of $\sim$
$620 \mbox{ kms}^{-1}$. The greatest compression occurs at the
front of MC, with the maximum $B=32 \mbox{ nT}$, $VB_z=19 \mbox{
mVm}^{-1}$, the minimum $B_z= 31 \mbox{ nT}$. However the highly
compressed magnetic field is northward and makes no contribution
to geomagnetic storm. This MC-shock event results in the
geoeffectiveness with a value of $Dst= -107 \mbox{ nT}$.

\begin{figure}
   \includegraphics[scale=0.7]{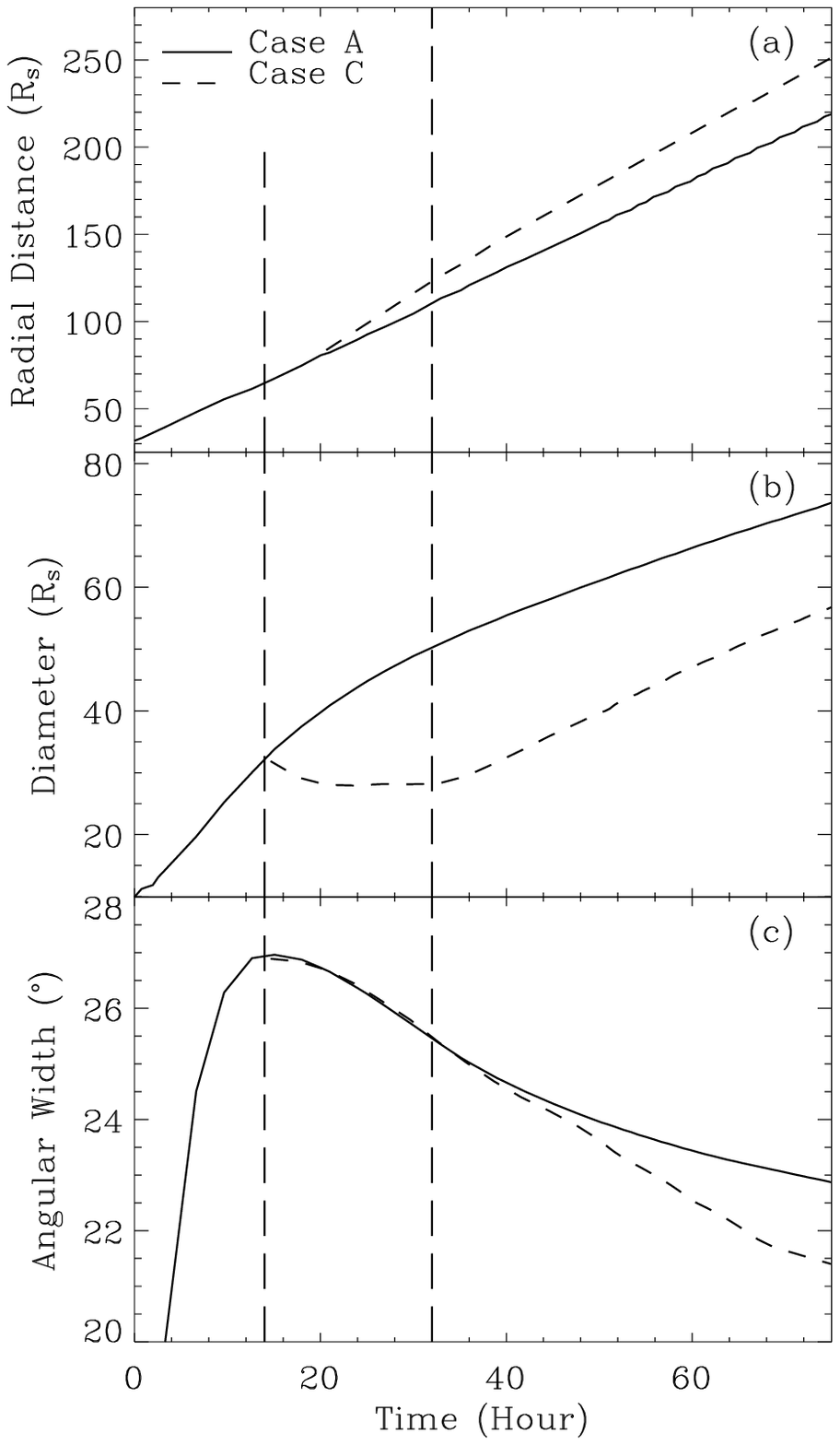}
\caption{} \label{multi-geometry}
\end{figure}

The comparison among Cases A and C about time-dependent parameters
is shown in Figure~\ref{multi-geometry}. Firstly, heliospheric
distance of MC core in Case A depends nearly linearly on time, as
shown in Figure~\ref{multi-geometry}(a). The solar-terrestrial
transporting speed of MC core is approximately $486 \mbox{
kms}^{-1}$. It suggests that an individual MC moves at a constant
speed through IP medium, consistent with relevant simulations
\citep{Vandas1995,Vandas1996b,Groth2000,Manchester2004a}.
Meanwhile one can see that MC core in Case C is compressed by the
shock, beginning from 20 hours. Secondly, MC boundary is not a
exactly circle due to overall force balance at MC-ambient flow
interface. MC diameter, defined as the radial distance difference
between its inner and outer boundaries along HCS, is still used to
quantify the size of MC. One can see that MC diameter increases
monotonically to $73 R_s$ at 1AU with an asymptotic radial
expansion speed $93.7 \mbox{ kms}^{-1}$ after 55 hours in Case A,
as indicated by Figure~\ref{multi-geometry}(b). However, the
angular width of MC in Case A behaves differently, as shown in
Figure~\ref{multi-geometry}(c). It undergoes an initially rapid
expansion from $15^\circ$ to $27^\circ$, then gradually recovers
to $23^\circ$ at 1AU. Physical interpretation for the variance of
MC width is as follows: (1) The MC abruptly expands at initial
stage because its inherent magnetic field is overwhelming over
that of the ambient solar wind. (2) It contracts gradually
afterwards while propagating in IP medium as its magnetic field
decreases faster than that of IMF. Meanwhile, the diameter and
width of MC in Case C is compressed by the shock, as indicated in
Figure~\ref{multi-geometry}(b) and (c). Finally, the relationship
between magnetic field magnitude in MC core and the time in Case A
is also sought for the power $\zeta$ in $B \propto t^{-1/\zeta}$.
$\zeta$ is about 0.76 in our model, consistent with relevant
results \citep{Vandas1995,Vandas1996b}. The expansion of
individual MC is pronounced from Figure~\ref{multi-geometry}, even
on the condition of the adiabatic process $\gamma=5/3$. Hence our
simulation is in favor of the idea that $\gamma<1$, proposed by
\citet{Osherovich1993a,Osherovich1993b,Osherovich1995}, may not be
a strict limitation for IP MC expansion
\citep{Vandas1996b,Vandas1996c,Vandas2000,Skoug2000,Vandas2003}.

Moreover the disturbance of speed enhancement just downstream of
incidental shock front can not completely propagate into MC
medium. After the shock front enters MC medium, the remaining high
speed flow follows right after the inner boundary of preceding MC
all the time, as seen from Figure \ref{Fig:CaseC}(a)-(c), which
can also be seen in the relevant simulation (Figure 3 in
\citet{Vandas1997a}). The MC is highly compressed by the
overtaking shock, as shown in Figure \ref{multi-geometry}(b). The
MC diameter decreases monotonically during shock passage in MC
medium (14 $\sim$ 32 hrs). It then recovers gradually when shock
penetrates and emerges from the MC ($> 32$ hrs). Compared to
relevant simulation (Figure 7(b) in \cite{Lugaz2005}), the
behavior of MC diameter in our simulation differs only after shock
emergence from MC medium. Because the forward shock in
\cite{Lugaz2005} is driven by the following MC, the diameter of
preceding MC remains constant by the compression of following MC
body when the shock propagates at the front of preceding MC. With
the push from above-mentioned high speed flow
(Figure~\ref{Fig:CaseC}(a)-(c)) instead of following MC body, the
MC diameter can not be completely recovered to that in
corresponding individual MC event (Case A) after the passage of
fast shock.

\section{Geoeffectiveness Studies}\label{Sec:Geoeffect}

\begin{figure}
   \includegraphics[width=20pc]{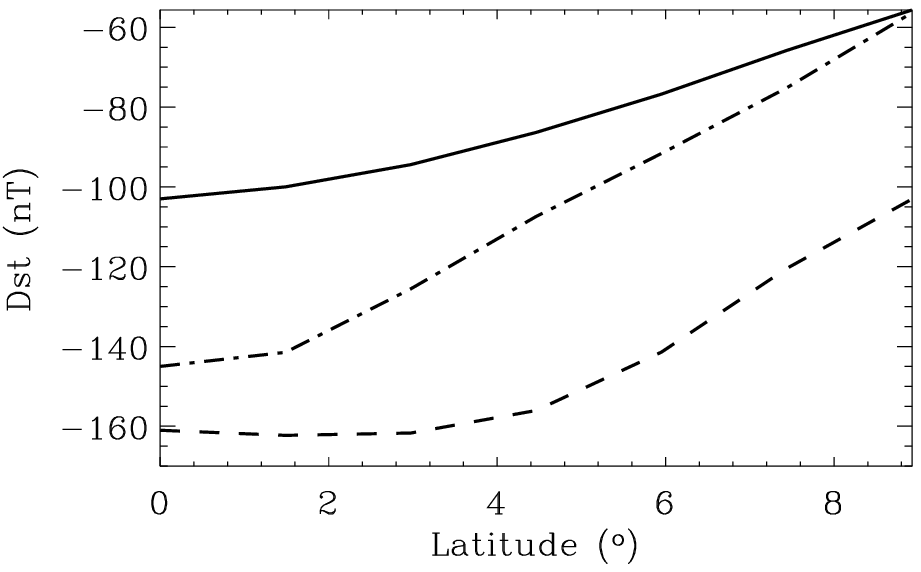}
\caption{} \label{multi-lat}
\end{figure}

Near-HCS latitudinal dependence of the $Dst$ index is plotted in
Figure~\ref{multi-lat}. The Geomagnetic storm has been obviously
aggravated by the shock overtaking the MC. The minimum $Dst$ is
found to be -103 nT in case A, -162 nT in case B, and -145 nT in
case C. In Particular, the latitudinal distribution of $Dst$ in
Case B is nearly constant over Lat. $=-4^\circ \sim 4^\circ$. On
the one hand, the southward passing magnetic flux decreases
steadily away from the equator because the MC propagates along the
HCS. On the other hand, the morphology of the shock front is a
concave astride the heliospheric equator when the shock penetrates
into the MC and has just begun to change, compared with a
well-established smooth arc in Case C. The greatest compression
occurs outside the equator. Two factors are balanced over a
certain latitudinal width, thus resulting in the above-mentioned
level distribution of $Dst$ in Case B.

It is found from Cases B and C that the geoeffectiveness of
MC-shock compound is undermined when shock penetrates completely
through MC. To further study the dependence of $Dst$ value on the
penetration depth of shock overtaking MC, a set of numerical
simulations with different duration between the emergence times of
MC and shock are carried out. Seventeen cases are run with
$t_{s0}=$ 3, 6, 10, 15, 20, 23, 26, 29, 32, 35, 38, 41, 44, 46,
48, 50, and 60 hours respectively.

\begin{figure}
    \includegraphics[width=20pc]{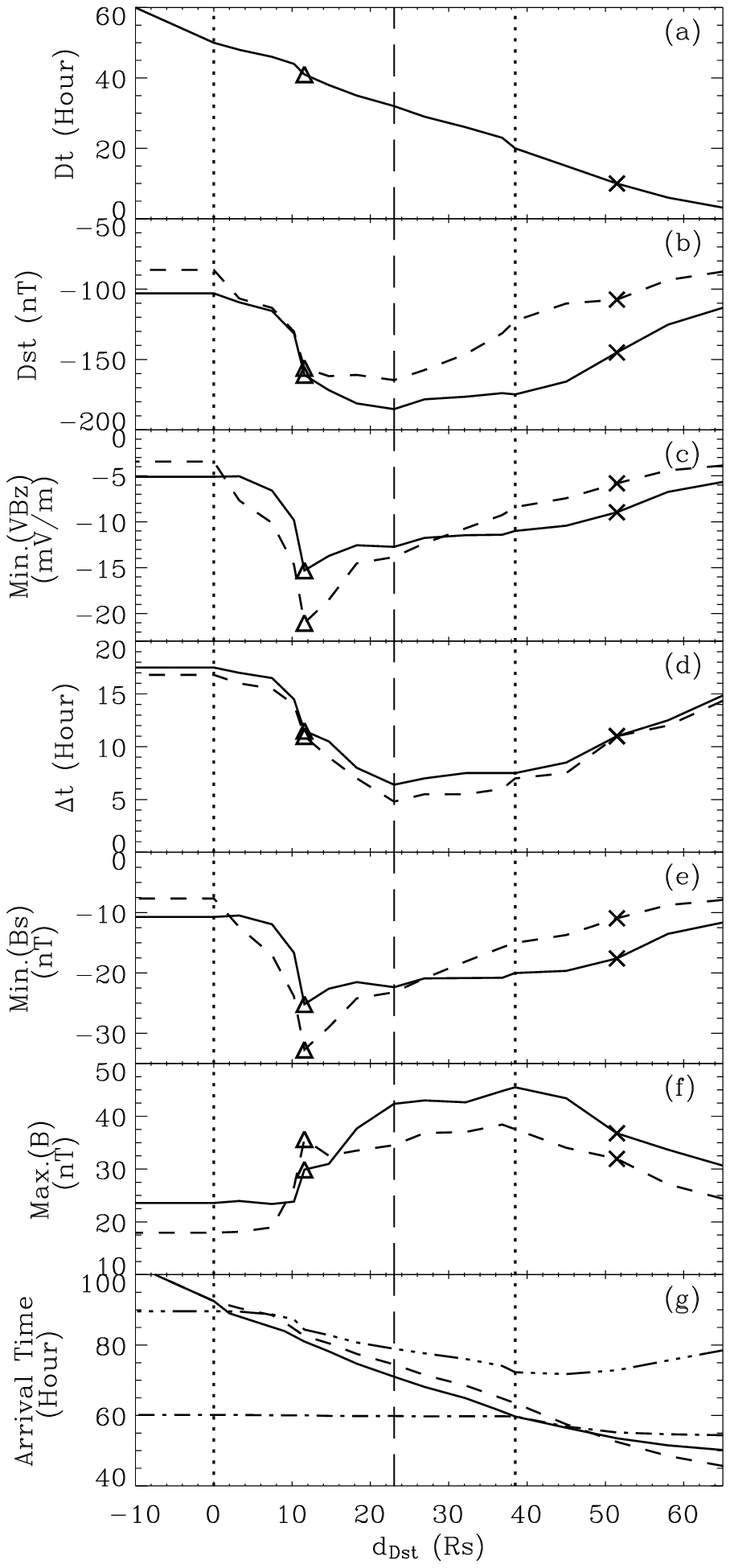}
\caption{} \label{depth3}
\end{figure}

By introducing a variable $d_{Dst}$, referring to the radial
distance along heliospheric equator between shock front and the
inner boundary of MC, we study the time-dependent data at L1
simultaneously recorded by two hypothetic spacecraft locating
along Lat. $=0^\circ$ and $4.5^\circ$ respectively. The
geoeffectiveness of MC-shock compound is described by $Dst$ as an
integral effect and minimum dawn-dusk electric field $VB_z$ as an
instant effect. Synthetical analyses on some crucial parameters
are given in Figure~\ref{depth3}, where the three vertical
delimiting lines (dotted, dashed and dotted) from left to right
correspond to the cases of shock encountering the tail, the core
and the front of MC at L1, respectively. From top to bottom are
plotted (a) the duration between the emergences of MC and shock
from the inner boundary, noted by $Dt$, (b) the $Dst$ index, (c)
the minimum of dawn-dusk electric field $VB_z$, noted by
$~\mbox{Min.}(VB_z)$, (d) the interval between the commencement of
$VB_z < -0.5~\mbox{ mV/m}$ and the corresponding $Dst$ minimum,
noted by $\Delta t$, (e) the minimum of $B_s$, noted by
$\mbox{Min.}(B_s)$, (f) the maximum of magnetic field magnitude
$\mbox{Max.}(B)$, and (g) the arrival times of the MC and the
shock along the equator, respectively. The solid and dashed lines
in Figure~\ref{depth3}(b)-(f) correspond to the hypothetic
satellites located at $\mbox{Lat.=}0^\circ$ and $4.5^\circ$,
respectively. It can be seen that MC and shock interact with each
other and merge into a complex compound structure when $Dt<$ 50
hours. The shock penetrates into the preceding MC more deeply with
less duration between MC and shock emergences. $\mbox{Min.}(B_s)$
and $\mbox{Min.}(VB_z)$ decline sharply, especially along Lat. =
$4.5^\circ$ as $d_{Dst}$ increases from 0 to $11R_s$.
Geoeffectiveness responses along Lat. = $4.5^\circ$ more
dramatically due to the concave front of incidental shock. This
results in almost the same $Dst$ value over Lat. $=0^\circ \sim
4.5^\circ$ for $d_{Dst} = 8 \sim 11 R_s$. Obviously, the minimum
$Dst$ with -185~nT along Lat. = $0^\circ$ and -165~nT along Lat. =
$4.5^\circ$ is obtained when the shock front just approaches MC
core at L1, corresponding to $d_{Dst}=23 R_s$, as indicated by
vertical dashed line of Figure~\ref{depth3}. Moreover,
$\mbox{Max.}(B)$ remains constant during $d_{Dst} = 0 \sim 7 R_s$,
because increasing magnetic field magnitude at compressed region
of MC is not yet comparable to that at MC core. The minimum of
$\mbox{Min.}(B_s)$ and $\mbox{Min.}(VB_z)$ are obtained at a
certain position in the rear part of MC. When the shock front
exceeds the MC core and compresses its anterior part with $23 R_s
< d_{Dst} < 38.5 R_s$, in which the magnetic field is northward,
$Dst$ recovers in different slopes along two latitudes. It
recovers gradually from -185 to -175 nT along Lat. = $0^\circ$ but
more rapidly from -165 to - 122 nT along Lat. = $4.5^\circ$. When
$D t < 20$ hours, the shock penetrates and propagates completely
through the MC before L1. The region of dramatic interaction
shifts from MC body to MC-driven sheath. As a result, magnetic
field tension of MC body overcomes the grip of post-shock total
pressure and $Dst$ continues to recover monotonically as $D t$
decreases. In addition, the minimum of $\Delta t$ (5 hours) and MC
passage interval (13 hours) correspond to $d_{Dst}=23$ and $38.5
R_s$ respectively. In contrast with -103 nT along Lat. = $0^\circ$
in corresponding individual MC event, $Dst$ reaches its minimum
-185 nT along the same latitude with 80\% increment in intensity
when the shock front advances into MC core. Moreover, the shock
transport time in MC-shock cases is shortened within $0 R_s <
d_{Dst} < 48 R_s$ in contrast with that in the corresponding
individual shock event, as indicated in Figure~\ref{depth3}(g).
The shortened time is 3.8 hours in maximum, corresponding to
$d_{Dst}=38.5 R_s$. When the shock propagates from IP medium to MC
medium, enhanced local magnetosonic speed and decreased bulk flow
speed upstream shock front coexist. The joint effect of these two
factors determines whether the shock is faster or slower in MC
medium. Hence the propagation speed of incidental shock influenced
by the MC depends on the interval between their commencements.

\section{Concluding Remarks and Discussions}\label{Sec:Conclusion}

Using a 2.5D ideal MHD numerical model, MC-shock interaction and
its geoeffectiveness are investigated for better understanding of
the IP ``shock overtaking MC" events
\citep{Wang2003b,Berdichevsky2005}. Our compound numerical
algorithm is capable of capturing sharp shock front, ensuring the
absence of magnetic monopole, guaranteeing the conservation of
axial and toroidal magnetic flux of magnetic rope, and so on. The
simulations reveal dynamic characteristics of IP MC-shock
interaction and their associated geoeffectiveness in some aspects.

Firstly, numerical simulation is carried out on an individual MC with its inherent magnetic field
overwhelming over that in the ambient flow. Characteristics of the specific MC propagation through IP space
are summarized as follows: (1) The MC core propagates with a nearly constant speed; (2) Its diameter
expands rapidly at initial stage. It then expands with a slower asymptotic speed; (3) Its angular width
also expands rapidly at initial stage, but gradually contracts afterwards. Moreover, the characteristics of
an MC, such as strong magnetic field, smooth rotation of magnetic field, low proton temperature, low plasma
$\beta$, and so on, are quite in agreement with the observations.

Secondly, numerical simulation is conducted to model MC-shock
interaction. A strong fast shock centered at HCS emerges from the
inner boundary to pursue the preceding MC. It is found that the
compression and rotation of magnetic field serve as an efficient
mechanism to cause large geomagnetic storm. The fast shock
initially catches up with the preceding MC. It then penetrates
through the MC and finally merges with the MC-driven shock into a
stronger compound shock. When the fast shock propagates through IP
space, its front is characterized with a central concave shape in
the equator; When it enters the preceding MC, its front evolves
into a purely arc shape. The morphology of shock front is
determined by the local medium. After the shock front enters MC
medium, the remaining high speed flow just downstream of
incidental shock front can not completely enter the preceding MC,
and it just follows behind the MC all the time. The MC is highly
compressed by the overtaking shock. The solar-terrestrial
transport time of incidental shock relates closely to the duration
between the emergences of MC and itself.

Lastly, the associated geoeffectiveness is studied based on
numerical simulations. In contrast with the corresponding
individual MC event, MC-shock interaction results in a largest
geomagnetic storm with 80\% increment in terms of $Dst$. Based on
an analytical solution for the process of shock propagation from
the inner boundary to the center of MC, \citet{Wang2003c}
suggested that the maximum geomagnetic storm be caused by shock
penetrating MC at a certain depth, and the stronger the incident
shock is, the deeper is the position. Meanwhile, the incidental
shock in our simulation is very strong and the results show that
the maximum geomagnetic storm occurs when the shock front
encounters MC core. Our numerical model agrees to some extent with
that by \citet{Wang2003c}. Furthermore, the high speed flow right
after the tail of MC boundary in our simulation mentioned
previously might be responsible for the minor difference of shock
penetration depth between the two models regarding the maximum
geomagnetic storm.

One can see that the compressed sheath field ahead of MC in our simulations is generally northward and,
hence, contribute little to geoeffectiveness (Figure~\ref{mc.L1}, \ref{mc-shock-B.L1},
\ref{mc-shock-C.L1}). If both MC helicity and ambient IMF orientation are reversed, the magnetic field
within MC-driven sheath and front part of MC will be directed southward and, hence, will be responsible for
geomagnetic storm. Some of qualitative results compared to that discussed above can be straightforwardly
conceived as follows: (1) Only when a shock propagates into the front of an MC does the shock exert its
effect on geoeffectiveness; (2) A shock losses its energy and momentum heavily during its propagation
through the rear part of an MC, so that it has relatively weaker influence on the geoeffectiveness by
penetrating the preceding MC. Moreover, if an incidental shock is not strong enough, it may be dissipated
quickly even in the rear part of an MC. Detailed quantitative investigation should resort to numerical
simulation. This interesting topic will be addressed in near future.

\begin{acknowledgments}
We express our heartfelt thanks to the referees for their constructive comments. This work was supported by
the National Natural Science Foundation of China (40274050, 40404014, 40336052, 40525014 and 40574063), and
the Chinese Academy of Sciences (startup fund). M. Xiong was also supported by Innovative Fund of
University of Science and Technology of China for Graduate Students (KD2005030).
\end{acknowledgments}


%
%
%

\begin{table}
\caption{Physical parameters of ambient solar wind at the bottom $(25R_s)$ and at Lagrangian point $L1$
$(213 R_s)$}\label{Tab1}
\begin{tabular}{|l|l|c|c|}
\hline
Variable  &  Description  &$25 R_s$  & $213 R_s$ \\
\hline
$N_p (~\mbox{cm}^{-3})$ & proton number density & 550 & 8  \\
$v_r  ~\mbox{(km/s)}$   & radial speed          & 375 & 452 \\
$B  ~\mbox{(nT)}$   & magnetic field strength& 400 & 6.4\\
$\beta$  & thermal to magnetic pressure ratio & 0.23 & 0.93 \\
$T_p (~\mbox{10}^5K)$ & proton Temperature & 9.6 &0.7 \\
$c_f ~\mbox{(km/s)}$ & radial fast characteristic speed & 372 & 61\\
\hline
\end{tabular}
\end{table}

\section*{Figure Captions}
\begin{description}
\item[Figure 1] One snapshot of a typical MC near L1 for Case A.
(a) Magnetic field magnitude $B$, (b) radial flow speed $v_r$, and
(c) proton beta $\beta_p$ are illustrated with two additional
radial profiles along Lat. = $0^\circ$ and $4.5^\circ$
respectively. Note: radial profile of $B$ is plotted by
subtracting initial ambient value $B|_{t=0}$. The white solid line
in each image denotes the boundary of MC. The difference of
magnetic flux function $\Delta \psi$ between adjoining magnetic
field lines in and out of the MC are $5.9 \times 10^{12}$ and $7.9
\times 10^{12} \mbox{ Wb}$. Solid and dashed lines at each profile
denote the core and boundary of MC. [\textit{See the electronic
edition of the Journal for a color version of this figure.}]

\item[Figure 2] The in-situ measurements along Lat.$=4.5^\circ$ by a hypothetic spacecraft at L1 for Case
A. Magnetic field magnitude $B$, elevation of magnetic field $\Theta$, radial flow speed $v_r$, proton beta
$\beta_p$, proton temperature $T_p$, calculated dawn-dusk electric field $VB_z$, and $Dst$ index are
plotted in stacked panels. Solid and dashed vertical lines denote the center and boundary of MC.

\item[Figure 3] The evolution of shock overtaking MC for Case B,
with (a)-(c) magnetic field magnitude $B$, (d)-(f) radial flow
speed $v_r$, and (g)-(i) radial characteristic speed of fast mode
$c_f$. Below each image are two attached radial profiles along
Lat.$=0^\circ$ and $4.5^\circ$. Note: spatial profile of $B$ is
plotted by subtracting initial ambient value $B|_{t=0}$.
[\textit{See the electronic edition of the Journal for a color
version of this figure.}]

\item[Figure 4] The in-situ measurements along Lat. $=4.5^\circ$ by a hypothetic spacecraft at L1 for Case B.

\item[Figure 5] The evolution of shock overtaking MC for Case C
with radial flow speed $v_r$. Only part of domain is plotted to
highlight MC in (a) and (b). [\textit{See the electronic edition
of the Journal for a color version of this figure.}]

\item[Figure 6] The in-situ measurements along Lat.$=4.5^\circ$ by a hypothetic spacecraft at L1 for Case C.

\item[Figure 7] The time dependence of MC parameters: (a) radial distance of MC core, (b) MC diameter, (c) MC
angular width. The solid and dashed lines denote individual MC event (Case A) and MC-shock event (Case C).
Two vertical dashed lines denote when the shock front arrives at the rear and front of MC.

\item[Figure 8] The comparison of latitudinal distribution of $Dst$ index among individual MC event (Case A) and
MC-shock events (Cases B and C). The solid, dashed, and dashed-dotted lines denote Case A, B, C
respectively.

\item[Figure 9] The parameter variances of MC-related geoeffectiveness as a function of $d_{Dst}$. Here $d_{Dst}$
refers to radial distance between shock front and inner MC boundary along heliospheric equator. From left
to right, three vertical lines (1st dotted, dashed, 2nd dotted) denote the critical situations of shock
just reaching the tail, the core, and the front of preceding MC at L1 respectively. The mark $\Delta$ and
$\times$ denote corresponding numerical results of Case B and C. (a) $Dt$, the duration between the
emergences of MC and shock from the inner boundary, (b) $Dst$ index, (c) $~\mbox{Min.} (VB_z)$, the minimum
of dawn-dusk electric field $VB_z$, (d) $\Delta t$, the interval between the commencement of
$VB_z<-0.5~\mbox{ mV/m}$ and the corresponding $Dst$ minimum, (e) $~\mbox{Min.}(Bs)$, the minimum of
southward magnetic component, (f) $~\mbox{Max.}(B)$, the maximum of magnetic magnitude and (g) arrival
times of the MC and the shock along the equator, respectively. Solid and dashed lines in (b) to (f)
correspond to observations along $~\mbox{Lat.=}0^\circ$ and $4.5^\circ$. Arrival times of the outer and
inner boundaries of MC, as well as that of incidental shock in MC-shock event and corresponding individual
shock event are indicated by dashed-dotted and dashed-dotted-dotted, as well as solid and dashed lines in
(g) respectively.

\end{description}

\newpage

\clearpage

\end{article}

\end{document}